\documentstyle [a4,psfig,12pt] {article}

\begin{document}
\begin{center}
{\Large\bf Electromagnetic Dipole Response as a Test  of the\\
 $^{\bf 11}$Li g.s. Structure and the n-$^{\bf 9}$Li Interaction}\\
{}~ \\
{}~ \\
{\bf Russian-Nordic-British Theory (RNBT) collaboration}\\
{}~ \\
B.V. Danilin, \\
{\em The Kurchatov Institute, 123182 Moscow, Russia}\\
M.V. Zhukov, \\
{\em Department of Physics, Chalmers University, G\"{o}teborg, Sweden}\\
 J.S.Vaagen,\footnote{Permanent address: Institute of Physics,
University of Bergen, Norway} \\
{\em NORDITA, DK-2100 Copenhagen \O, Denmark} \\
I.J.Thompson, \\
{\em Department of Physics, University of Surrey, Guildford GU2 5XH, U.K.}\\
J.M.Bang\\
{\em The Niels Bohr Institute, DK-2100 Copenhagen \O, Denmark} \\
\end{center}

\begin{abstract}
The electric dipole response of the halo nucleus $^{11}$Li is calculated in a
hyperspherical three-body formulation, and is  studied as a function of the
interaction employed for $n-^9$Li to reflect the Pauli principle.  Strength
concentrations at lower energies are  found  but no narrow  resonances. Only
one possible scenario of $^{11}$Li structure is in close correspondence with
MSU and RIKEN experimental data.
 \end{abstract}

\centerline{For submission to {\em Physics Letters}}

\newpage

Very recently, the experimental data for the electric dipole strength
distribution $dB({\cal E}1,E)/dE$ have become available from MSU
\cite{MSU} for the two-neutron halo nucleus $^{11}$Li at excitation
energies  $E \leq 2$ MeV. The Michigan fragmentation experiment for
$^{11}$Li  + Pb  was carried out at $28A$ MeV. While previous data on
inclusive variables such as geometrical properties and momentum
distributions for individual break-up fragments have proven unable to
discriminate between quite different wave functions for the bound state of
$^{11}$Li \cite{bang92,zhuk91,DEN}, we will show that the dipole response
function sheds some light on this issue. We remind the reader about the
peculiar 3-body Borromean structure of $^{11}$Li: since there is only one
bound state (weakly bound at about 300 keV) and none of the binary
subsystems are able to form bound states, the asymptotic conditions of the
problem are pure 3-body asymptotics.

Fig. 1 shows the deconvoluted experimental Michigan data which peaks at
around 0.6 MeV, with energy being measured from the break-up threshold.
The same peak position was reported\cite{shim92} for the RIKEN experiment
with the same reaction at $42A$ MeV. Fig. 1 also shows previous
theoretical results, that of the simple point-dineutron cluster model of
Bertulani and Bauer (BB)\cite{bert91,bert92},  and the more realistic
model of Esbensen and Bertsch (ESB)\cite{esb92}. We notice that the
calculations, which nearly coincide, are shifted to low energies compared
with the data. In the BB cluster model the energy dependence is simply
$E_s^{1/2} E^{3/2}/(E+E_s)^4$, where $E_s\approx 0.3$ MeV is the
separation energy of the two halo neutrons. This distribution peaks at $E
= \frac{3}{5}E_s$, giving a linear dependence on the separation energy.
The Green's function method of ESB employs a $\delta$-type $nn$
interaction.

This letter investigates three limiting cases of plausible dynamics for
the halo neutrons of $^{11}$Li, corresponding to three tentative
prescriptions for treating the Pauli Principle within a strict 3-body
formulation. Expansion of hyperspherical harmonics (HH) is used \cite{HH},
a natural procedure for this Borromean system.

We have recently \cite{HE6,dan91,dan92} and with considerable success,
carried out within the same HH framework bound state and continuum
calculations for $^6$He, also a Borromean nucleus, which shares many
properties with $^{11}$Li. Contrary to $^{11}$Li, however, sufficient
information is available on the binary $n$-core interaction, and this
makes physically reliable calculations possible. Both the $0^+$ ground
state and the known $2^+$ resonance at an excitation
energy of 1.8 MeV were reproduced by our calculations. However, although
these calculations gave strength concentrations at low continuum energies,
they showed no narrow resonances neither in the phase shifts nor in the
calculated strength functions for electric dipole
excitations of $^6$He.
A parallel study of $^{11}$Li is hampered by lack of information on the
$n-^9$Li channel, although calculations can technically be carried
out\cite{zhuk91,bang92}.

The dipole response has the form
\begin{eqnarray*}
B({\cal E}1;0^+_{\rm g.s.} \rightarrow 1^-(E)) =
 \mid \langle 1^-(E) || T_{1}^{{\cal E}} || 0^+_{\rm g.s.} \rangle\mid^2,
\end{eqnarray*}
where
$     T_{1M}^{{\cal E}} = \sum _{i} e_{i}r_{i}Y_{1M}(\hat{r}_{i})$
is the corresponding intrinsic 3-body cluster dipole operator
with coordinates referred to the c.m. of $^{11}$Li, and hence free from
spurious c.m. motion. Note that the operator explores both the $1^-$
structure of the continuum as well as the structure of the $0^+$ ground
state. (We leave the assumed $^9$Li(3/2$^-$) core out of all our
calculations and discussion.) With previous experience from a similar
calculation for $^6$He \cite{dan92}, we expect that replacing
six-dimensional plane waves in the $n-n$ and $(nn)-^9$Li coordinates by
correlated states from strict 3-body continuum calculations will shift the
continuum strength to lower energies. Fig. 2 shows that this expectation is
borne out in all cases. Notice that leaving out continuum correlations for
the ESB case moves the peak into the position of the data, but that it now
falls significantly below the data in strength.


We follow \cite{bang92,thomp92} and consider three choices, referred to as
the H, WS-P and WS-F scenarios.  For all cases considered, the $nn$
interaction is that of Gogny-Pires-de Tourreil (GPT)\cite{GPT} and
includes repulsion at small distances, as well as spin-orbit and tensor
forces.  The Pauli principle is included by introducing repulsive `Pauli
cores' from the required combination(s) of central and spin-orbit forces.
As discussed in previous papers \cite{zhuk91,bang92}, all scenarios give
a binding energy for $^{11}$Li of $\sim 0.3$ MeV, i.e. within experimental
error, and give \cite{DEN} very  similar single-particle densities. As in
the $^6$He case, investigations of the three-body scattering amplitudes
\cite{dansur} showed no narrow dipole resonances states
for any of the scenarios.
These three scenarios give however very different
ground-state wave functions, as exhibited in table 1.

In the WS-F (Woods-Saxon-Forbidden) scenario, the spin-orbit force is
assumed to play a leading role.  We assume that there is a $0p_{1/2}$
resonance in $n$ + $^9$Li scattering (\cite{RES9,BERLIN,RES9B}), with the
$0p_{3/2}$ existing as a bound state of $^{9}$Li (B.E.= 4.1 MeV).  The
$^{9}$Li core is then taken to have a full level of $0p_{3/2}$ neutrons,
and the Pauli principle was taken into  account by pure repulsive
potential in  the occupied $0s$ and $0p_{3/2}$ states of neutron-core
motion. The resulting $^{11}$Li wave function (WF) will be predominantly
$(0p_{1/2})^2$, which corresponds (for harmonic oscillator orbitals) to a
linear combination of 33\% $^1S_0$ and 67\% $~^3P_1$ states of
neutron-neutron motion.  This implies that the $^{11}$Li ground state has
a rather small probability of a di-neutron configuration, starting at 33\%
and increasing to 43\% (see table 1) when  correlations are included.

In the WS-P (with P for pairing) scenario, we follow \cite{HAY} by
taking the pairing rather than the spin-orbit forces as dominant in
$^{11}$Li. A Cohen-Kurath type of calculation for $^9$Li gives \cite{HAY}
a ground state that is 93\% of the symmetry [$f$] = [432], which can
couple {\em only} with the spatially symmetric ([$f$] = [2]) neutron pair
to form $^{11}$Li with a closed shell structure.  The valence neutrons are
now almost entirely in ($l_{nn} = 0$, $S = 0$)  relative motion, with the
$S=1$ configuration blocked by the core neutrons. The spin-orbit force
being damped in the interior so that it does not destroy the coherence of the
state which is approximately  $\sqrt{2/3}(0p_{3/2})^2 + \sqrt{1/3}
(0p_{1/2})^2$.  We include this effect in our model by having no
neutron-core spin-orbit forces, taking into account only the $S = 0$
configurations in the $0p$ shell. As in WS-F case we use a Pauli
repulsive $s$-wave core-$n$ interaction.

The third (H) scenario has no explicit Pauli terms, but simply employs the
shallow Gaussian potential of \cite{JJH,zhuk91}, which does not support any
occupied orbits. Thus  the mean fields for the halo and core neutrons
differ substantially. The halo neutrons may now be largely in $0s$ motion
relative to the core, and only partially blocked by Pauli orthogonality
with the core neutrons because the core and valence radii are different.
 This calculation also reproduces the binding energy and r.m.s. matter radius
of the $^{11}$Li ground state and gives the largest value of hyper-radial
moments $< \mid \rho^4 \mid >$ compared to $< \mid \rho^2 \mid >^2$ (see Table
1) as an expression of the softness of the $^{11}$Li halo system.

In all scenarios, the ground state of $^{11}$Li has a closed neutron
shell: $(0s)^2$ in the H case, and $(0s)^2(0p)^6$ in the WS-P and WS-F
cases. Thus, to form a $1^-$ excitation, a neutron will have to be excited
into the next shell of opposite parity,  implying an energy which is
$\hbar\omega$ $\approx$ 4 MeV  in a standard estimate for dipole
excitations.

For halo nuclei, however, the neutrons are very near threshold, and simple
estimates based on $\hbar\omega$ $\approx$ 4 MeV may no longer be correct.
Recently, for example, it was  argued in a self-consistent density-functional
method \cite{fay91}, and also in a two-body cluster calculation \cite{bert92},
that the centre of gravity of strength functions should be concentrated closer
to zero energy as the valence level approaches threshold. To answer this
question, we have calculated continuum distributions in a 3-body model, which
should include the effect of such shifts as a 3-body threshold feature.


In all scenarios, we obtain a satisfactory  convergence for the ${\cal
E}1$ strength in the energy region below 10 MeV . However, for energies
greater than about 4 or 5 MeV (the threshold for $^9$Li excitation) more
complicated mechanisms come into play, and our analysis assuming an inert
core is no longer physically valid.
We must also ensure convergence in the radial integrations.
Since the effective range (a product of two wave functions and a radial
operator) of the  dipole operator is $\sim~40$ fm,  to obtain the response
without erroneous contributions
from artificial low lying  additional structures, we have extended our
calculations out to 150--200 fm.

The results for continuum final states (using the same  3-body hamiltonian
as for the g.s.) are shown in Fig. 1, compared with the MSU data. We show
also the results (ESB) of the  Green function method\cite{esb92}, and the
distributions (BB) of the 2-body cluster model\cite{bert91}.  The WS-P and
WS-F distributions, as expected from simple estimates, are very broad.
The continuum RPA results\cite{bert90} are similar to the WS-P curve. Only
the H scenario resembles the MSU data.

The curves peak at different energies. The position of the BB peak has
already been discussed.  In the WS-P model, qualitatively speaking,  we
have a purely repulsive $s$-wave $n$-core  interaction, lifting up the
levels in the next shell, and in WS-F there is also a $p_{3/2}$ repulsion,
additionally pushing the levels to higher energies.
The curves also approach $E=0$ in different ways. In the 2-body
BB\cite{bert92} point dineutron cluster model an $E^{3/2}$ behaviour is
obtained for the low-energy ${\cal E}1$ response.  Since the g.s. WF is
concentrated in the asymptotic tail, it is possible to estimate this
quantity in the three-body case, which gives $\sim~E^{3}$ for the
threshold behaviour. Both models give $\sim E^{-5/2}$ when $E$ is much
larger than the binding energy.

In the calculation of continuum WFs the ground state core-$n$ potential
and realistic GPT $nn$ interaction were used. If we replace this $nn$
interaction by a simple central Gaussian (which however reproduces $n-n$
low energy phase shift data), all curves shown in figs. 1 \& 2 are changed
in the H case by less than the line width, and by only a few percent in
the WS cases.  Only the core-$n$  and central part of the $nn$ interaction
are decisive for the $^{11}$Li dynamics.


In our calculations of the electric dipole  response of $^{11}$Li with 3-body
wave functions for both the continuum and g.s.,
we summarise our results as follows:

 i) Although the  strength is concentrated at lower energies, resembling
resonant behaviour, no narrow resonances are found in the three-body
scattering amplitudes \cite{dansur}, implying a wide spreading of dipole
states into the continuum;

 ii) Our calculations support qualitatively the conclusion that the
enhancement of dipole strength at low energy is mainly due to the halo
structure of the g.s. That is, the kinematical (threshold effect)
enhancement is due to the large moments of the mass distribution, and to
the proximity of the g.s. to the three-body core + $n$ + $n$ decay
threshold.

 iii) ${\cal E}1$ transitions which play the main role in electromagnetic
dissociation  are very sensitive to: 1) the structure of the ground state
structure and to  a lesser extent of the continuum; 2) the method of
treating the Pauli principle; and (not addressed here) possible mechanisms for
the $^{11}$Li dynamics during the reaction process.

%
 iv) Only the H-scenario is in a good qualitative agreement with the
experimental ${\cal E}1$ deconvoluted response. It should be noted, however,
that in all three cases  the Pauli principle was taken into account
in an approximate way. The question of how to handle the Pauli principle in
halo
nuclei is thus still open, and one we are still persuing.

We will return to the nuclear monopole mode and the quadrupole excitation
in  a larger paper, where also sum rules are discussed in more detail.

We wish to acknowledge support from NORDITA and the University of Bergen,
where much of this work was carried out. U.K. support from SERC grants
GR/H53648 and GR/H53556 is acknowledged. Helpful discussions with H. Schulz
and experimental results communicated by him are also appreciated.

\newpage
{\bf FIGURES}

\centerline{\psfig{figure=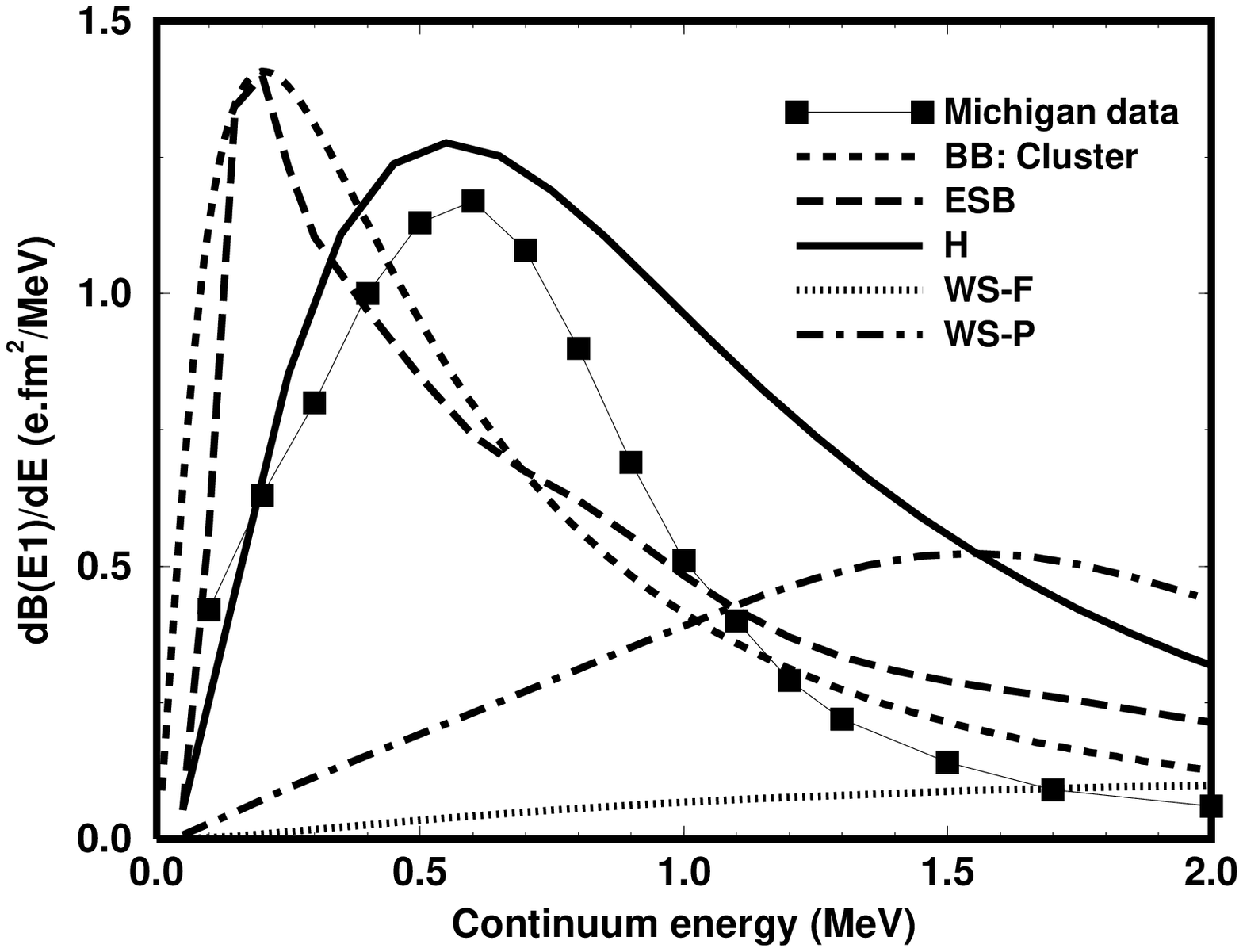,height=7cm}}
\begin{quotation}
{\em Figure 1:}
Electric dipole response for $^{11}$Li from the  MSU experiment \cite{MSU}
(squares), with
the Greens function \cite{esb92} (ESB, long dashed) and cluster model
\cite{bert92} (BB, short dashed) calculations.
The curves H (solid), WS-F (dotted),  WS-P (dot-dashed) are
the predictions from the current 3-body scenarios.
 \end{quotation}

\centerline{\psfig{figure=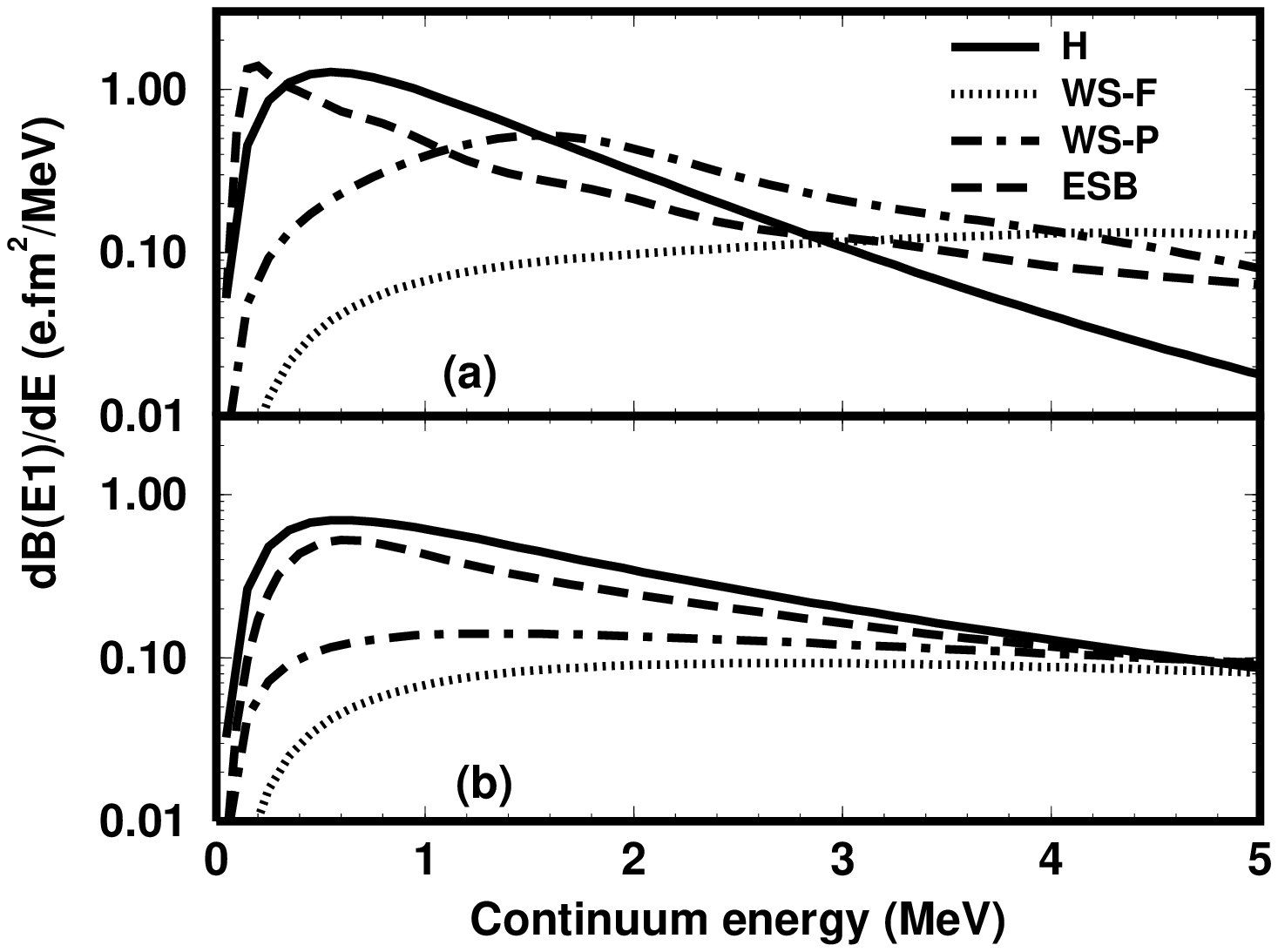,height=9cm}}
\begin{quotation}
{\em Figure 2:} Electric dipole response with correlated (a) and uncorrelated
(b)
continuum wave functions, for $^{11}$Li in the  H, WS-F \& WS-P scenarios,
and the ESB  calculation \cite{esb92}.
 \end{quotation}

\newpage
\bigskip
\subsection*{Tables}

\begin{table}[tbh]
\begin{center}
\begin{tabular}{|ccccc|l|l|l|}
\hline
 K & L & S & $l_{nn}$ & $l_{(nn)c}$ & WS-F & WS-P & H \\
\hline
 0 & 0 & 0 & 0 & 0 &   1.17 \%&   2.69 \% & 96.25 \% \\
 2 & 0 & 0 & 0 & 0 & 40.93& 94.55 &  2.48\\
 2 & 1 & 1 & 1 & 1 & 54.21& 0    & 0\\
\hline
 4 & 0 & 0 & 0 & 0 & 1.30& 1.44 & 0.36\\
 4 & 1 & 1 & 1 & 1 & 0.19& 0    & 0\\
 4 & 0 & 0 & 2 & 2 & 1.88& 0.29 & 0.66\\
\hline
$\geq6$ &   & all &  &  & 0.31& 1.02 & 0.22\\
\hline
 $E_{s}$, MeV & & & &          & 0.332& 0.248& 0.295\\
\hline
$ r_{mat}$, fm  & & & &          &   2.97&   3.00&   3.31\\
\hline
$< \mid \rho^2 \mid >,$ fm$^2$& & & & &   48.3&   48.0&   76.3\\
\hline
$< \mid \rho^4 \mid >,$ fm$^4$ & & & & & 4129  & 4936  &16135  \\
\hline
$\hat{ E_{{\cal E}1}},$ MeV & & & & &    4.7&    2.8&    1.6\\
\hline
\end{tabular}
\end{center}
\caption{Partial norms, matter radii, binding energies for $^{11}$Li g.s.
in WS-P,  WS-F  and H scenarios, and mean energies for  ${\cal E}1$
responses. $K$ is the hypermoment quantum number, while $l_{nn}$  and
$l_{(nn)c}$ are orbital angular momenta in the $n-n$ and $(nn)-^9$Li
degrees of freedom, and couple up to $L$.}

\end{table}


\end{document}